\documentclass[twocolumn,prd,superscriptaddress]{revtex4-2}
\usepackage{graphicx}
\usepackage{physics}
\usepackage{color}
\usepackage{amsmath}
\usepackage{amsfonts}
\usepackage{xspace}
\usepackage{array}
\usepackage{makecell}
\usepackage{subfigure}
\usepackage{ulem}
\usepackage{epstopdf}
\usepackage[dvipsnames]{xcolor}
\usepackage[breaklinks,colorlinks,linkcolor=blue,citecolor=blue,urlcolor=blue]{hyperref}
\usepackage{ulem}
\def\avg(#1){\langle#1\rangle}

\def\be{\begin{equation}}
	\def\ee{\end{equation}}
\def\bea{\begin{eqnarray}}
	\def\eea{\end{eqnarray}}

\def\nn{\nonumber}

\usepackage{ulem}

\begin{document}	
	
	\title{Enhancing  analogue Unruh effect via superradiance in a cylindrical cavity}%
	\author{Hong-Tao Zheng}
	\author{Xiang-Fa Zhou}
	\email{xfzhou@ustc.edu.cn}
	\author{Guang-Can Guo}
	\author{Zheng-Wei Zhou}
	\email{zwzhou@ustc.edu.cn}
	\affiliation{CAS Key Lab of Quantum Information and Anhui Center for fundamental sciences in theoretical physics, University of Science and Technology of China, Hefei, 230026, China}
	\affiliation{Synergetic Innovation Center of Quantum Information and Quantum Physics, University of Science and Technology of China, Hefei, 230026, China}
	\affiliation{ Hefei National Laboratory, University of Science and Technology of China, Hefei 230088, China}
	
	\date{\today}	
	
	\begin{abstract}
		We propose a  scheme to detect the Unruh effect in a circularly rotated Unruh-DeWitt detector enclosed within a cylindrical cavity. This technique relies on the enhanced  atomic spontaneous emission rate related to the counter-rotating coupling between the detector and massless scalar fields. Our analysis demonstrates that the integration of a cylindrical cavity, coherent light excitation, and multi-atom super-radiation significantly enhances the signal strength, as the radiation rate associated with the standard rotating-wave coupling can be greatly suppressed within the cavity. Compared to linear acceleration, circular motion can significantly reduce the atomic acceleration path length, leading to increased detection efficiency and lower experimental difficulty. Our method  provides a novel avenue for exploring relativistic effects on a compact, tabletop platform. 
	\end{abstract}

	\maketitle
	
	\textit{Introduction.}
	General relativity and quantum mechanics are recognized as the two cornerstones of modern physics. Although it still remains challenging to integrate them into a more general theory,  there are many frameworks that can bridge these two fields, including Hawking radiation \cite{HAWKING1974} and the Unruh effect \cite{PhysRevD.14.870}, among others.
	In particular, the Unruh effect has garnered sustained attention in recent years due to its profound physical implications. For example, a potential solution to the perplexing black hole information paradox has been proposed through the generalized Unruh effect \cite{PhysRevD.107.056014}. Quantum information, and quantum teleportation between Alice and Bob in different frames have also been widely discussed \cite{PhysRevLett.91.180404, PhysRevLett.95.120404,
		PhysRevA.80.032315, PhysRevA.106.032432}.  Entanglement harvesting utilizing Unruh-DeWitt detectors (UDDs) has been explored in various studies \cite{PhysRevA.70.012112, Salton_2015, PhysRevD.105.085012, PhysRevD.102.065013}.
	Moreover, the intricate relationship between the Unruh effect and other physical phenomena, such as Cherenkov radiation, time refraction, dynamical Casimir effect, and the radiation reaction has also been extensively discussed \cite{ PhysRevLett.126.063603, MENDONCA20085621, PhysRevD.65.065015, Johnson2005}.  All these explorations provide us with new insights into the possible unified framework of physics.

	However, the experimental confirmation of the Unruh effect remains elusive, as it is still a huge technical challenge to simultaneously achieve uniform accelerated motion for particles and detect the Unruh emission within the same  setup. Despite this, there are two principal directions being pursued to observe this effect in the laboratory.
	On the one hand, indirect experimental detection methods have been proposed and implemented in various systems. For instance, in cold atomic systems, the first UDD has been proposed to observe the Gibbons-Hawking effect, in the context of the de Sitter analog of the Unruh effect\cite{PhysRevLett.91.240407, PhysRevD.69.064021}. Then, the Unruh effect has also been explored by simulating the curved spacetime metric or Hamiltonian \cite{PhysRevA.95.013627, 10.21468/SciPostPhys.5.6.061,PhysRevLett.101.110402, Hu2019, PhysRevResearch.2.042009,PhysRevLett.125.213603, PhysRevA.102.033506}.
	Furthermore, the detection and simulation of the Unruh effect using Berry phase \cite{PhysRevLett.107.131301, PhysRevA.85.032105, PhysRevD.106.036013, PhysRevD.106.045011, PhysRevLett.129.160401} or tapered optical fibers \cite{SMOLYANINOV20085861, Ge:21} have been considered as viable alternatives. 
	%These approaches hold the potential to provide crucial insights into the nature of the Unruh effect and bring us closer to a direct experimental verification.
	On the other hand, the direct experimental detection schemes considering the possibility of using accelerated electrons to detect the Unruh effect have been widely discussed\cite{BELL1983131, PhysRevD.52.3466, PhysRevD.58.084027, UNRUH1998163, BELL1987488, PhysRevLett.61.2113, PhysRevD.105.096034}. Other schemes involving the employment of UDD as a quantum detection device have also been proposed as a promising avenue to verify the effect \cite{PhysRevLett.91.243004, PhysRevA.74.023807, PhysRevLett.125.241301, PhysRevLett.128.163603, PhysRevLett.129.111303}. (We also recommend that readers refer to the review paper \cite{RevModPhys.80.787} and their references for details.)

	Physically, the UDD can be characterized as a two-level quantum system that is coupled to a scalar field \cite{RevModPhys.80.787}. Initially, the UDD is set in its ground state and moves along a worldline within a vacuum environment.  The  inertial UDD remains  unresponsive due to the invariance of the vacuum state in this case. However, when the UDD experiences a linear acceleration, it can be  excited and emits a photon due to the presence of counter-rotating effect in the laboratory frame. This emission can also be understood that,  the field is no longer in a vacuum state for the detector.
	Proposals for observing the response of the UDD due to non-inertial motion have been extensively discussed in Refs \cite{PhysRevLett.129.111303, PhysRevLett.125.241301}. However, both of these studies primarily focus on the emission rate induced by rotating-wave coupling. 
	Within the framework of optical cavity systems, another promising approach has been extensively discussed for detecting the probability associated with acceleration-induced counter-rotating transitions \cite{PhysRevLett.91.243004,PhysRevLett.91.243004}.  In this case,  a sufficiently large linear acceleration and cavity size are required, both of which are  still difficult to achieve in laboratory.
	% This approach necessitates a significant level of linear acceleration and a cavity of considerable size, both of which remain challenging to realize in a laboratory setting. 
	In Refs \cite{wang2021}, a novel strategy is discussed that involves "many" oscillating UDDs  within a microwave cavity. This approach aims to utilize a Dicke superradiance-like vacuum amplification effect  to achieve linear growth of signal intensity with the number of atoms. However, the requirement of a significantly large number of detectors makes the experimental control still extremely challenging.

	In this article, we discuss the coupling between the UDD  and a massless scalar field that satisfies the Dirichlet boundary condition within a cylindrical cavity.  As the UDD rotates uniformly around the axis of the cavity,  it experiences an acceleration of constant magnitude in varying directions. 
	We show that the probability of spontaneous emission resulting from counter-rotating terms exceeds that observed under typical rotating-wave coupling conditions, which enables us to effectively detect relativistic effects induced by the detector's motion.
	Furthermore,  by collective coupling of $N$ atoms to a specific frequency light field, we demonstrate that the combination of the superradiance effect and the mode selection of the  cavity greatly enhances the signal intensity, which scales proportionally to $N^2 |\alpha|^2$ with $|\alpha|^2$ the intensity of the laser beams. This significantly reduces the experimental difficulty,  and thus makes it more feasible to achieve on existing experimental platforms.
	Our work provides a new approach to detect relativistic effects induced by motion on laboratory platforms.

\begin{figure}[t]
		\includegraphics[width=0.45\textwidth]{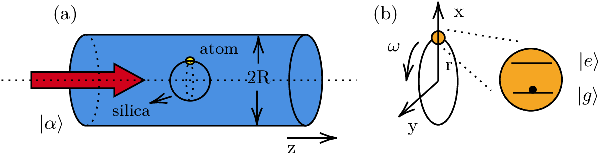}
		\caption{(a) Illustration of the coupling between atomic rotation and the vacuum field (or coherent state $\ket{\alpha}$) inside a cylindrical  cavity with radius $R$; (b) The atom rotates with the orbital radius $r$ and angular velocity $\omega$ in the $xy$ plane. }
		\label{fig-setup}
	\end{figure}

	\textit{The model. }  
	We start by considering a two-level atom inside a cylindrical cavity of radius $R$, which interacts with a massless scalar field $\Phi(x)$, as schematically shown in Fig. \ref{fig-setup}.  The total Hamiltonian of the system can be expressed as 
	\begin{equation}
		H=\frac{\Omega}{2}\sigma_z+\int dk \omega_k a^{\dagger}_ka_k+g\sigma_x\Phi(\textbf{x})
	\end{equation}
	where $\Omega$ is the energy splitting, $\sigma_z=\ket{e}\bra{e}-\ket{g}\bra{g}$  and $\sigma_x=\ket{e}\bra{g}+\ket{g}\bra{e}$ are the Pauli matrices with $\ket{g},\ket{e}$ the ground and excited states of the atom respectively. $\omega_k$ is the frequency of the photon with the annihilation and creation operators defined by $a_k$ and $a^{\dagger}_k$. The last term describes the interaction between the atom and the field with the strength $g$, which takes the similar form of usual dipole approximation as the light-matter interactions discussed in quantum optics. 
	In the interaction picture, the interaction part can be rewritten as 
	\begin{equation}		\label{eq_Hint}
		H_{int}=g\sigma_x(\tau)\Phi(\textbf{x},t)=g(e^{i\Omega\tau}\sigma_{+}+h.c.)\Phi(\textbf{x},t)
	\end{equation}
	where $\tau$ is the proper time of the atom, $\sigma_+=\sigma_{-}^{\dagger}$ satisfies $\sigma_+=|e\rangle\langle g|$. The field operator reads $\Phi(\textbf{x},t)=\int dk a_k\phi(\textbf{x},t)+a_k^{\dagger}\phi^*(\textbf{x},t)$, where $\phi_k(\textbf{x},t)$ is the mode solution for scalar fields. The interaction $H_{int}$ simultaneously includes both the rotating-wave term and the counter-rotating term. In typical quantum optical systems, the counter-rotating term oscillates rapidly which usually results in a negligible average effect. However, in our case, the motion of the particles can greatly enhance the counter-rotating term effect, making it a crucial element for detecting the Unruh effect.

	The counter-rotating term can induce a counter-intuitive transition of the atom from  the ground state $\ket{g}$ to the excited state $\ket{e}$, emitting a photon carrying the momentum $k$. In the first-order perturbation theory, the probability of this type of spontaneous emission is given by
	\begin{equation}
		\begin{split}
			P_{cr}(k)&=\left|\int d\tau \bra{1_k,e}H_{int}\ket{0,g} \right|^2=g^2|I_{cr,k}|^2\\
			&=g^2\int_{-\infty}^{\infty}d\tau d\tau' e^{-i\Omega(\tau-\tau')}|\bra{1_k}\Phi(\textbf{x},t)\ket{0}|^2,         
		\end{split}
		\label{eq: P(k)}
	\end{equation}
	where$\ket{0,g}$ and $\ket{1_k,e}$ represent the composite states of the laser fields and the atom, $I_{cr,k}=\int_{-\infty}^{\infty}d\tau e^{i\Omega \tau}\phi_k^*(\textbf{x},t)$ is the frequency component of the mode $\phi_k^*(\textbf{x},t)$ defined by the proper time $\tau$ of the atom. If we sum over all the momenta of the photons emitted from the atom, and concentrate on the probability of exciting the atom, we can integrate Eqs. (\ref{eq: P(k)}) over $k$ and obtain
	\begin{eqnarray}	
		P_{cr}=g^2\int_{-\infty}^{\infty}d\tau d\tau' e^{-i\Omega(\tau-\tau')}W(x,x'),
	\end{eqnarray}
	where $x=(\textbf{x},t), x'=(\textbf{x}',t')$, 
	and $W(x,x')=\bra{0}\Phi(\textbf{x},t)\Phi(\textbf{x}',t')\ket{0}$ is the Wightman function\cite{RevModPhys.80.787}. Defining $\tau_{+}=\frac{\tau+\tau'}{2}$ and $\tau_{-}=\tau-\tau'$, the spontaneous emission rate of counter-rotating terms can be expressed as 
	\begin{equation}
		\Gamma_{cr}(\Omega)=\frac{P_{cr}}{T_t}=g^2\int_{-\infty}^{\infty}d\tau_{-} e^{-i\Omega\tau_{-}}W(x,x'). \label{eq:GammaCR}
	\end{equation}
	where $T_t=\int_0^T d\tau_{+}$ is the detection time of UDD. Similarly, for rotating terms, where the detector makes a transition from $\ket{e}$ to $\ket{g}$ and emits a photon with the energy-momentum vector $(\omega_k,k)$, we just need to replace $\Omega$ with $-\Omega$ for a reverse transition of the atom compared to counter-rotating terms. The relevant spontaneous emission rate can also be approximated as $\Gamma_{r}(\Omega)=\Gamma_{cr}(-\Omega)$.
	%\begin{equation}
	%		\begin{split}
		%			P_{r}(k)&=\left|\int d\tau \bra{1_k,g}H_{int}\ket{0,e}\right|^2=g^2|I_{r,k}|^2\\
		%			&=g^2\int_{-\infty}^{\infty}d\tau d\tau' e^{i\Omega(\tau-\tau')}|\bra{1_k}\Phi(\textbf{x},t)\ket{0}|^2
		%		\end{split}
	%       \label{eq: P_R(k)}
	%\end{equation}
	%with $I_{r,k}=\int_{-\infty}^{\infty}d\tau e^{-i\Omega\tau}\phi^{*}(\textbf{x},t)$, which takes almost the same form as Eqs.(\ref{eq: P(k)}) expect that we just need to replace $\Omega$ with $-\Omega$. Following the same routeline, we can rewrite the emission rate induced by rotating-wave terms as 
	% \begin{equation}
		% 		\Gamma_{r}(\Omega)=\Gamma_{cr}(-\Omega)=g^2\int_{-\infty}^{\infty}d\tau_{-} e^{i\Omega\tau_{-}}W(x,x').  \label{eq:GammaR}
		% \end{equation}    Equations (\ref{eq:GammaCR}) and (\ref{eq:GammaR}) 
	These discussions indicate  that both of the two spontaneous emission rates $\Gamma_{r}$ and $\Gamma_{cr}$ depend closely on the explicit form of the field model $\Phi(\textbf{x},t)$, which comprises the basic starting points for latter discussions.
	
	\textit{Unruh effect  induced by an atom inside a cylindrical cavity with uniform circular motion.} 
	Due to the presence of cavity boundaries, only specific frequencies and optical field modes can exist within the cavity. %As we have outlined above, compared to linear acceleration, uniform circular motion can greatly reduce the difficulty of implementation in real experimental system. This makes it convenient for us to explore the relativistic Unruh effects caused by motion in the laboratory. 
	For a cylindrical cavity satisfying the Dirichlet boundary condition $\Phi(r=R,\theta,z,t)=0$,  the massless scalar field $\Phi(\textbf{x},t)$ inside the cavity can then be expressed as the superposition of basic Bessel beams, and is given by\cite{PhysRevLett.129.111303}
	\begin{equation}
		\begin{split}
			\Phi(\textbf{x},t)=&\frac{1}{2\pi R}\sum_{p=-\infty}^{\infty}\sum_{q=1}^{\infty}\frac{B_{p}(\xi_{pq}r/R)}{B_{|p|+1}(\xi_{pq})}\int_{-\infty}^{\infty}\frac{dk_z}{\sqrt{\omega_k}}\\
			&*(a_ke^{-i\omega_kt+ip\theta+ik_zz}+H.c.),
		\end{split}
		\label{Phi}
	\end{equation}
	where $\xi_{pq}$  is the $q$-th zero of Bessel function $B_{p}$ and satisfies $\xi_{pq}>p$, $\omega_k^2=c^2k_z^2+\frac{c^2\xi_{pq}^2}{R^2}$ is the energy of output photons with the momentum $k$. 
	For an atom undergoing uniform circular motion around the axis inside the cavity, the corresponding space-time coordinates of its trajectory in the lab frame can be represented as $(t,x,y,z)=(\gamma \tau, r\cos(\gamma\omega\tau),r\sin(\gamma\omega\tau),z_0)$,
	where $\omega$ is the angular velocity of the atom with the orbital radius $r$, and $\gamma=1/\sqrt{1-r^2\omega^2/c^2}$ is the Lorentz factor. The relevant spontaneous emission rates read: 
	\begin{eqnarray}\label{eq:gammaRCR}
		\Gamma_{r(cr)}=\frac{g^2}{ \pi R^2}\sum_{p=-\infty}^{\infty}\sum_{q=1}^{\infty}C_{pq}\frac{\Theta(p\omega\pm\frac{\Omega}{\gamma}-\frac{c\xi_{pq}}{R})}{\sqrt{(p\omega\pm\frac{\Omega}{\gamma})^2-(\frac{c\xi_{pq}}{R})^2}},
	\end{eqnarray}
	where $C_{pq}=\frac{B^2_{p}(\xi_{pq} r/R)}{B^2_{|p|+1}(\xi_{pq})}$ depends on the relative ratio of $r$ and $R$, and $\Theta(x)$ is the Heaviside step function with $\Gamma_r$ and $\Gamma_{cr}$ corresponding to $p\omega+\frac{\Omega}{\gamma}$ and $p\omega-\frac{\Omega}{\gamma}$, respectively(See Appendix A for details).

	Equation (\ref{eq:gammaRCR}) indicates that the spontaneous emission rates have local maximum values at the resonant points defined by   $\gamma p\omega=\gamma\frac{c\xi_{pq}}{R} \pm\Omega$.
	For normal emission driven by the rotating-wave term, the atom relaxes from the excited state to emit a photon satisfying $\gamma p\omega=\gamma\omega_k -\Omega$.  Physically,  the presence of cavity boundaries provides a lower bound for the allowed modes $\omega_k$ inside the cavity.
	If the energy gap $\Omega$ of the atom is much lesser than the frequency of the lowest eigen-mode inside the cavity,  $\Gamma_r$ can be greatly suppressed when $\omega=0$ since the cavity modes are off resonant with the atom. Mathematically, this  corresponds to the case with $\Omega<\frac{c\xi_{pq}}{R}<\omega_k$ in the system.  It is noteworthy that this is impossible in a borderless environment, as $\Omega=\omega_k$ can always be satisfied by increasing the radius $R$ of the cavity. In this latter case,  $C_{0q}(\xi_{0q}r/R)$ approaches $1$ as $R\rightarrow \infty$, while other terms tend to $0$ in Eq.(\ref{eq:gammaRCR}) (For example, we have $C_{01}/C_{11}\approx 10^{11}$ when $r/R=10^{-6}$). Therefore, the normal spontaneous emission is always dominated in the absence of the cavity.

	For $\Gamma_{cr}$ related to the counter-rotating process, the atom is excited from the ground state and emits a photon, which means $\gamma p\omega=\gamma\omega_k+\Omega$ and $p\omega>\omega_k\ge\frac{c\xi_{pq}}{R}$ in this case. 
	Meanwhile, since $\xi_{pq}>p$, we have $\omega R>c$. This indicates that for the cavity mode with $\omega_k$, the velocity of the atom is faster than the local phase velocity of the light in the cross-section perpendicular to the z-direction, leading to the occurrence of counter-rotating  coupling.  This mechanism is analogous to the standard  Cherenkov radiance\cite{PhysRevLett.126.063603}, and similar discussions can also be found in Ref. \cite{PhysRevD.53.4382}. 
	Physically, such cascade excitation process induces a continuously transfer of energy into the system by rotating atoms, akin to the linear acceleration case\cite{PhysRevLett.126.063603}.
	Compared with usual rotating-wave coupling, we conclude that the counter-rotating coupling requires a larger angular velocity $\omega$ to fulfill the resonant conditions for the same $\omega_k$.

	%%%%%%%%%%%%%%%%%%%%%%%%%%%%%%%%%%%%%%%%%%
	\begin{figure}[t]
		\includegraphics[width=0.4\textwidth]{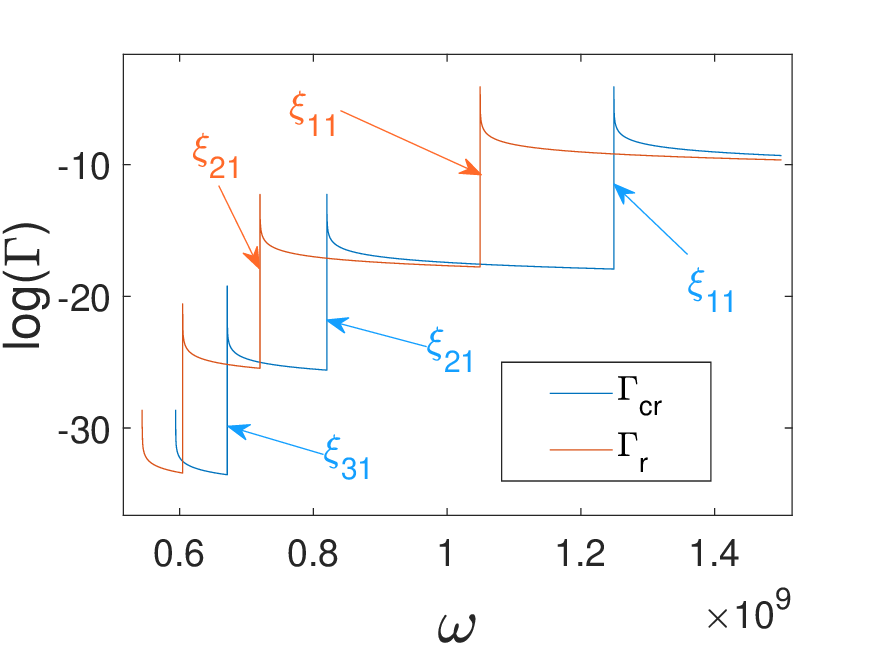}
		\caption{Spontaneous emission rates $\Gamma_{cr}$ (blue) and $\Gamma_{r}$ (red) over the angular velocity $\omega$ of the atom  with the orbital radius $r=0.01{\rm m}$. Other parameters are as follows: $R=1{\rm m}, \Omega=100{\rm MHz}, g=10{\rm kHz}, r=0.01{\rm m}$, and $q=1$.}
		\label{fig1}
	\end{figure}
	%%%%%%%%%%%%%%%%%%%%%%%%%%%%%%%%%%%%%%%%%%%%%
	
	Figure (\ref{fig1}) shows how $\Gamma_{cr}$ and $\Gamma_r$ change as functions of the angular velocity $\omega$ for other fixed parameters. 
	We can check that, around each resonant point, there is always a range of $\omega$ where $\Gamma_{cr}$   is consistently higher than $\Gamma_r$. 	
	For the specific parameters shown in Fig. \ref{fig1}, the spontaneous emission rate is about $\Gamma_{cr}\simeq e^{-4}=0.019{\rm s}^{-1}$ at the resonant point $(p,q)=(1,1)$, indicating the atom will be excited after about $50$ seconds. The acceleration is estimated to be $\omega^2 r\simeq 1.56*10^{16}{\rm m}/{\rm s}^2$, and is far smaller than that of the linear acceleration, which is estimated to be on the order of $10^{26}{\rm m}/{\rm s}^2$ \cite{RevModPhys.80.787}. %However, it is still hard to set up the experimental conditions, so we put forward two ways to increase the spontaneous emission rate.
	However, the implementation of such experimental conditions is still currently challenging, which prompts us to further increase the rate of radiation. %This is also the focus of subsequent discussions.

	\textit{Enhancing the emission rates using Dicke-like superradiance.}
	One of the simplest methods to enhance the emission rate is to increase the number of atoms. Specifically, when atoms are close enough, they can collectively couple with cavity modes,  greatly increasing the rate of radiation. This phenomenon is known as Dicke superradiance \cite{PhysRev.93.99}. 
	The Hamiltonian of the Dicke-like model can be expressed as 
	\begin{equation}
		H=\Omega J_z+\sum_k\omega_k a^{\dagger}_k a_k+g(J_++J_-)\Phi(\textbf{x})
	\end{equation}
	where $\Omega$ is the gap of each detector. $J_z=\frac{1}{2}\sum_{k=1}^{N}\sigma_z^{k},  J_{\pm}=\sum_{k=1}^{N}\sigma_{\pm}^k$ and $J^2=(J_+J_-+J_-J_+)/2+J_z^2$ are the collective operators of $N$ spin-$1/2$ atoms. The collective states of these atoms can be represented as the common eigenstates  $\ket{j,m}$ of operators $\{J^2,J_z\}$ and satisfy $J^2\ket{j,m}=j(j+1)\ket{j,m}$ and $J_z\ket{j,m}=m\ket{j,m}$. In the interaction picture, the interaction part can be rewritten as 
	\begin{equation}		
		H_{int}=g(e^{i\Omega\tau}J_{+}+h.c.)\Phi(\textbf{x},t)
	\end{equation}
	which is similar to Eq. (\ref{eq_Hint}). Here, we set the total spin of atomic states as $j=\frac{N}{2}$. Thus, for an initial Dicke state of atoms $\ket{j,m}$, the above model indicates that the relevant spontaneous emission rates $\Gamma^D_{r(cr)}$ for rotating-wave (counter-rotating) terms can be obtained as 
	\begin{eqnarray}
		\Gamma^D_{r(cr)}=(j\pm m)(j\mp m+1)\Gamma_{r(cr)}. %\nonumber \\
		%\Gamma^D_{cr}&=&|\bra{j,m+1}J_+\ket{j,m}|^2\Gamma_{cr}\nonumber \\&=&(j+m+1)(j-m)\Gamma_{cr}. \nonumber
	\end{eqnarray}
	If all atoms are in the ground (or excited) state, i.e., $m=\mp\frac{N}{2}$, we have $\Gamma^D_{cr}=N\Gamma_{cr}(\mbox{or } 0)$ and $\Gamma^D_{r}=0(\mbox{or } N\Gamma_{r})$ respectively.  Both the emission rates are  simply increased by a factor of $N$.  This is expected, as the number of atoms has now multiplied by $N$ compared to the original count. However, when half of the atoms are excited, namely $m=0$, both relevant emission rates are enhanced by a factor of $\frac{N}{2}(\frac{N}{2}+1)$.   
	If we consider a scenario with $10^6$ atoms, both $\Gamma^D_{r}$ and $\Gamma^D_{cr}$ are then increased by 12 orders of magnitude compared to the case of a single atom. To achieve the same emission rates, the circular acceleration of the atoms can be reduced to $1.56\times 10^{12}{\rm m}/{\rm s}^2$.

	It is worth noting that in most proposals of the Unruh effect, atoms are assumed to be in their ground state initially, rendering the effect of rotating-wave coupling negligible. However, during superradiation, both $\Gamma_{cr}^D$ and $\Gamma_{r}^D$ increase simultaneously when $m=0$. The counter-rotating coupling effects become apparent only when $\Gamma_{cr}^D/\Gamma_{r}^D > 1$, which occurs near the resonance condition $\gamma p\omega =\gamma\omega_k+\Omega$, as shown in Fig. (\ref{fig1}).  In addition, the mode selection properties of the cavity enable the adjustment of the ratio $\Gamma_{cr}^D/\Gamma_{r}^D$ within a broad range  by introducing laser coupling at an appropriate frequency. As a result, the signal of the counter-rotating coupling can be more easily detected, thereby further reducing the experimental difficulty.
	% and therefore convenient to observe under this condition. 
	% However, the atoms will transition to ground state rapidly as $m=0$ due to $\Gamma_{cr}^D/\Gamma_{r}^D\gg 1$ in most systems. Thus enhancing Unruh effect by superradiance seems not work in this case. Fortunately, this difficulty can be overcome.}
% \textcolor{blue}{On the one hand, Figure (\ref{fig1}) show that around each resonance point, there is always a range of $\omega$ where $\Gamma_{cr}$  is consistently higher than $\Gamma_r$ in our system, which has been discussed before.} On the other hand, the relative ratio of $\Gamma_{cr}/\Gamma_r$ can also be increased using coherent laser beams. 
Physically, this can be achieved by appropriately choosing the angular velocity $\omega$ and $\omega_k$ so that only $\Gamma_{cr}$ is enhanced, while $\Gamma_r$ is still strictly suppressed.
For general coherent states input of the cavity mode $k$, the stimulated emission rate for a single atom in the ground state reads  (See Appendixes B and C for details)
\begin{eqnarray}
	\Gamma^{\alpha}_{cr,k}&=&\frac{1}{\pi T_t}\int d^2\beta_k |g\int d\tau\bra{\beta_k,e}\Phi(\textbf{x},t)\sigma_x(\tau)\ket{\alpha_k,g}|^2  \nonumber\\
	&=&\frac{g^2}{\pi T_t}\int d^2\beta_k e^{-|\alpha_k-\beta_k|^2}|\beta_k^*I_{cr,k}+\alpha_k I^{*}_{r,k}|^2
\end{eqnarray}
with $\ket{\alpha_k}$ and $\ket{\beta_k}$ representing the coherent states of the cavity mode $k$. 
% $T_t=\int_0^T d\tau_{+}$ is the detection time of UDD. 
Since $I_{cr,k}\sim \delta(\Omega+\gamma\omega_k-\gamma p\omega)$, if we  fix the angular velocity $\omega$ and choose the frequency of input pulse at resonant point defined by  $\Omega+\gamma\omega_k=\gamma p\omega$ with $p=1$, then $I_{cr,k}$ reaches its local maximum. Meanwhile, since $I_{r,k}\sim \delta(-\Omega+\gamma\omega_k-\gamma p\omega)$, the coupling related to rotating-wave terms is off-resonant. 
%The emission rate $\Gamma_{cr,k}$ reaches its local maximum at resonant points defined by  $\Omega+\gamma\omega_k=m\gamma\omega$ with $m=1$.
%Hence we can fixed the angular velocity $\omega$ of atom and the frequency $\omega_k=\omega-\Omega/\gamma$ of the driving beams so that only $\Gamma_{cr,k}$ is greatly enhanced, while the emission for rotating-wave coupling is still strongly suppressed as $-\Omega+\gamma\omega_k=\gamma\omega$ cannot be satisfied in the same time, as shown in Fig. \ref{fig-laserrate}. 
In this case, we have  $|I_{cr,k}|^2\gg |I_{r,k}|^2 \sim 0$ and $			\Gamma^{\alpha}_{cr,k}\simeq (|\alpha_k|^2+1)\Gamma_{cr,k}$,
where $|\alpha_k|^2$ is the average input number of photons, and $\frac{g^2}{T_t}|I_{cr(r),k}|^2=\Gamma_{cr(r),k}$ represents the spontaneous emission rate in mode $k$ without input laser beams. Therefore, by choosing the frequency $\omega_k$ appropriately, we can ensure that $\Gamma_{cr,k}^{\alpha}$ is completely determined by counter-rotating coupling. We  emphasis that in Ref. \cite{PhysRevLett.128.163603}, a special designed  world-line of the atom is always need  to ensure $\Gamma_{cr,k}>\Gamma_{r,k}$. However, in our case, we only need to keep the angular velocity of the atom constant. 

The above discussion can also be generalized to the case with multiple atoms. 
For the Dicke state, both the stimulated emission rates $\Gamma_{cr}^{\alpha,D}$ and $\Gamma_{r}^{\alpha,D}$ are significantly increased, while the relative ratio $\Gamma_{cr}^{\alpha,D}/\Gamma_{r}^{\alpha,D}$ can still remain high, thereby ensuring that $\Gamma_{cr}^{\alpha,D} \gg \Gamma_{r}^{\alpha,D}$ is satisfied (See Appendix C for details).
Furthermore, based on the result shown in  \cite{PhysRevApplied.18.064006}, we can also choose the initial state of atoms as $|\psi(0)\rangle=\ket{\frac{N}{2},\frac{N}{2}}$. After coupling to the coherent beam for a certain period of evolution, the system automatically evolves into a superradiant state(See Appendix D for details). This can then further reduce the experimental complexity.

% Specifically, for an initial state of atoms $|\psi(0)\rangle=\ket{\frac{N}{2},\frac{N}{2}}$ coupling to the coherent beam with the effective Rabi frequency $\Omega_R$, the mean value of the collective excitations $J_z(t)$ is given by $\langle J_z(t) \rangle=\frac{N}{2}\cos(\Omega_R t)$ \cite{PhysRevApplied.18.064006}. Hence the stimulated emission rates are significantly enhanced by superradiance when $t$ satisfies $\Omega_R t_c=\pi/2$ and $\langle J_z(t) \rangle|_{t_c}=0$.

\begin{figure}[t]
	\includegraphics[width=0.45\textwidth]{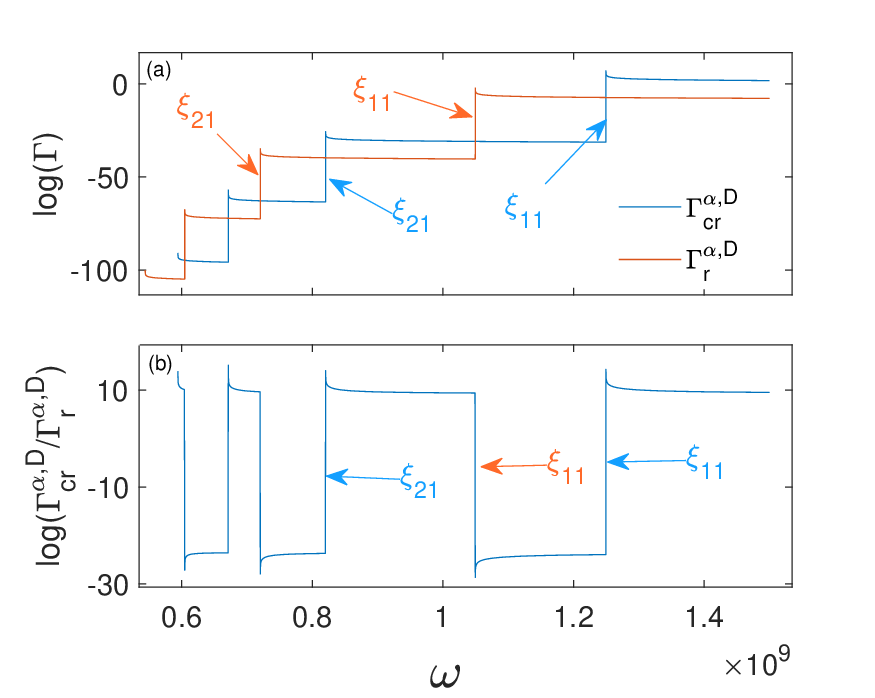}
	\caption{(a)  The stimulated emission rates $\Gamma^{\alpha,D}_{cr(r)}$ and (b) relative ratio $\Gamma^{\alpha,D}_{cr}/\Gamma^{\alpha,D}_{r}$ of $10^6$ atoms as functions of the angular velocity $\omega$ with the coherent state $|\alpha|^2=10000$ and orbital radius $r=50{\rm nm}$. In (a), the red line shows the effect of rotating-wave coupling,  and the blue line  corresponds to the effect of  counter-rotating terms; $\Gamma^{\alpha,D}_{cr}/\Gamma^{\alpha,D}_{r}$ exhibits abrupt discontinuity around the resonant points, as shown in (b). The specific settings for other parameters are as follows: $R=1{\rm m}, \Omega=100{\rm MHz}, g=10{\rm kHz}, q=1$.}  
	\label{fig-laserrate}
\end{figure}

For  $10^6$ atoms stimulated using coherent beams with $|\alpha|^2=10^4$,  the emission rate  $\Gamma_{cr}^{\alpha,D}$  is 16 orders of magnitude larger than $\Gamma_{cr}$. Fig. \ref{fig-laserrate} shows that the peak emission rate related to counter-rotating coupling is about $\Gamma_{cr}^{\alpha,D}\sim e^6=387{\rm s}^{-1}$ when the rotation speed reaches $\omega \simeq 1.25\times 10^9{\rm Hz}$, which means that on average, 387 atoms will be excited per second. Therefore,  detectable experimental results can be obtained in a few minutes. Experimentally, the rotation frequency of single $100 \rm{nm}$ particles exceeding $1{\rm GHz}$ has been observed in \cite{PhysRevLett.121.033602}. In our system, we can embed the atoms on the surface of the silica nanoparticle to achieve the high-speed rotation of the atoms, where the setup is similar to that in \cite{PhysRevLett.125.241301}. As for the coherent light $\omega_k=1.15\times10^9{\rm Hz}$, The wavelength is about $2{\rm m}$ at the radius direction and almost infinite along the z axis, which is so large that the phase differences between particles can be neglected. Thus it should be not hard to achieve the superradiance. Specifically, we would like to choose a cylindrical base rather than a spherical one due to the almost infinite wavelength along the z axis. All atomic detectors are supposed to be attached to the surface of the base with the same orbital radius to make the Dicke model valid. 
For instance, a cylindrical base with a radius of $100{\rm nm}$\cite{Jin:21} and a length of $500{\rm nm}$ along the z axis is large enough to achieve the superradiant effect of  $10^6$ atoms, considering a doping concentration of $10^{21}{\rm cm}^{-3}$\cite{WOS:000276554600023}. Alternatively, for a typical dilute atomic clound with densities $3\times 10^{15}{\rm cm}^{-3}$\cite{science.abi6153}, a cylindrical base with a radius of $300{\rm nm}$ and a length of $1{\rm mm}$ along the z axis can carry about $10^5$ atoms, which also achieves the same desired experimental results. %The main factor affecting the superradiant behavior is the orbital radius $r\simeq 50nm$ of atoms, which is related to the emission rate. All atoms are supposed to be embeded at the same orbital radius to make the Dicke model valid. Moreover, we would like to choose the cylindrical nanoparticle rather than the spherical one due to the almost infinite wavelength at the z axis. Specifically, the cylindrical nanoparticle with the radius 100nm\cite{Jin:21} and the length 500nm at the z axis is large enough to achieve $10^6$ atoms superradiance considering the doping concentration as $10^{21}cm^{-3}$\cite{WOS:000276554600023}. Alternatively, we might consider the ultracold atom clound with the densities $3\times 10^{15}cm^{-3}$\cite{science.abi6153}. In this case, we need a cylindrical nanopartcile with the radius 300nm and the length 1mm at the z axis to carry the $10^5$ atoms.(Since the emission rate is increasing rapidly following the radius growing, we need less atoms, $10^5$, to achieve the same experimental results. )}. 
Therefore, we believe it is conceivable that current technology may soon meet the requirements here, making it possible to be implemented in real platforms. The acceleration related to the peak emission rate can also be estimated as $\omega^2r \simeq 7.81\times 10^{10}{\rm m}/{\rm s}^2$, which is less than that for a single atom and is also significantly smaller than that in the linear accelerated case  \cite{RevModPhys.80.787}. 
Moreover, the relative ratio  $\Gamma^{\alpha,D}_{cr}/\Gamma^{\alpha,D}_r$ can exceed $10^4$, which greatly facilitates the detection of the counter-rotating wave effect. Specifically, for Dicke-like model discussed above, we will find that the ensemble emits photons constantly in the equilibrium case.  The equilibrium temperature $T$ of the ensemble (see Appendix E for details) can be defined by $\Gamma^{D}_{cr}/\Gamma^{D}_{r}=e^{-\beta \Omega}$ with  $\beta=1/T$. Therefore, the temperature can be negative as $\Gamma^{D}_{cr}>\Gamma^{D}_r$. That is very different from both the cases with linear acceleration\cite{Ben_Benjamin_2019}, and circular acceleration without borders\cite{PhysRevD.102.085006}.

\textit{Conclusion.} To summarize, by considering circularly accelerated atoms interacting with massless fields satisfying the Dirichlet boundary condition inside a cylindrical cavity, we propose a way to detect Unruh radiation by increasing the emission rate related to counter-rotating wave coupling.  With the help of collectively enhanced coupling using multiple atoms and stimulated superradiance via coherent laser beams, we can reduce circular acceleration to $7.81\times 10^{10}{\rm m}/{\rm s}^2$ at $\omega=1.25\times 10^9{\rm Hz}$\cite{PhysRevLett.121.033602} using typical experimental parameters. This is smaller than $1.56\times 10^{16}{\rm m}/{\rm s}^2$ discussed in previous studies and far smaller than that required by the linear-acceleration case. 
The equilibrium temperature of the atomic ensemble can be negative, indicating that atoms can continuously emit photons while maintaining their state unchanged. However, the effective temperature of the photons emitted from the atoms is more complex. Generally, we can not define  the effective temperature or the relevant Bose-Einstein distribution directly. However, it may be still possible to define an effective temperature under other conditions, which deserves a separate detailed discussion in future work.
The work can also be generalized to cover the dependence of detection efficiency due to the effect of dissipation and other varying boundary conditions. Our work thus provides a possible  approach for experimentally detecting relativistic motion effects on current physical platforms and may pave the way for further research on the influence of relativistic motion on quantum dynamics.

We thank Prof. X.-W. Luo for helpful discussions. This work was funded by National Natural Science Foundation of China (Grants No. 12474366, No. 11974334, and No. 11774332), and Innovation Program for Quantum Science and Technology (Grant No. 2021ZD0301200, ZD0202070101). XFZ also acknowledges support from CAS Project for Young Scientists in Basic Research (Grant No.YSBR-049).

\appendix
\vspace{0.2cm}

\section{\label{A} Spontaneous emission rate for a single atom interacted with vacuum }

For the massless scalar field $\Phi(\textbf{x},t)$ defined by Eq.(\ref{Phi}) in  the main text, the Wightman function can be obtained as
\begin{equation}
\begin{split}
	W(x,x')&=\bra{0}\Phi(\textbf{x},t)\Phi(\textbf{x}',t')\ket{0}\\
	&=\frac{1}{4\pi^2R^2}\sum_{p=-\infty}^{\infty}\sum_{q=1}^{\infty}C_{pq}\int_{-\infty}^{\infty}\frac{dk_z}{\omega_k}
	e^{-i\gamma\omega_k\tau_-+ip\gamma\omega\tau_-}
\end{split}
\end{equation}
where $\tau_-=\tau-\tau'$, and the trajectory of the atom in the lab frame can be represented as $(\textbf{x},t)=(r,\gamma\omega\tau,z_0,\gamma\tau)$ in the cylindrical coordinate system. The coefficient $C_{pq}=\frac{B^2_{p}(\xi_{pq} r/R)}{B^2_{|p|+1}(\xi_{pq})}$ depends on the relative ratio of $r$ and $R$. Therefore, the spontaneous emission rate of rotating-wave terms is then given by
\begin{equation}
\begin{split}
	\Gamma_{r}=&g^2\int_{-\infty}^{\infty}d\tau_{-} e^{i\Omega\tau_{-}}W(x,x')\\
	=&\frac{g^2}{4\pi^2R^2}\sum_{p=-\infty}^{\infty}\sum_{q=1}^{\infty}C_{pq}\int_{-\infty}^{\infty}\frac{dk_z}{\omega_k}
	\int_{-\infty}^{\infty}d\tau_{-} \\
	&e^{i\Omega\tau_{-}-i\gamma\omega_k\tau_-+ip\gamma\omega\tau_-}\\
	=&\frac{g^2}{\pi R^2}\sum_{p=-\infty}^{\infty}\sum_{q=1}^{\infty}C_{pq}\int_{\frac{c\xi_{pq}}{R}}^{\infty}\frac{d\omega_k\delta(\Omega-\gamma\omega_k+p\gamma\omega)}{\sqrt{\omega_k^2-\frac{c^2\xi_{pq}^2}{R^2}}}
	\\
	=&\frac{g^2}{ \pi R^2}\sum_{p=-\infty}^{\infty}\sum_{q=1}^{\infty}C_{pq}\frac{\Theta(p\omega+\frac{\Omega}{\gamma}-\frac{c\xi_{pq}}{R})}{\sqrt{(p\omega+\frac{\Omega}{\gamma})^2-(\frac{c\xi_{pq}}{R})^2}},
\end{split}
\end{equation}
where $\Theta(x)$ is the Heaviside step function. Similarly, we can get the spontaneous emission rate of counter-rotating terms by replacing $\Omega$ with $-\Omega$.

\section{Stimulated	emission rate for a single atom interacted with an input coherent state }
For general coherent states input $\alpha_k$ of the cavity mode $k$, the emission rate for a single atom from the ground state
$\ket{g}$ to the excited state $\ket{e}$ with an output coherent state $\ket{\beta_k}$ is given by 
\begin{equation}
\begin{split}
	\Gamma'_{\beta_k,k}&=\frac{1}{T_t}|g\int d\tau\bra{\beta_k,e}\Phi(\textbf{x},t)\sigma_x(\tau)\ket{\alpha_k,g}|^2\\
	&=\frac{1}{T_t}|g\int d\tau e^{i\Omega \tau}\bra{\beta_k,e}\Phi(\textbf{x},t)\sigma_+\ket{\alpha_k,g}|^2\\
	&=\frac{g^2}{T_t}|\int d\tau e^{i\Omega \tau}(\alpha_k\phi(\textbf{x},t)+\beta^*_k\phi^*(\textbf{x},t))|^2\\
	&=\frac{g^2}{T_t}e^{-|\alpha_k-\beta_k|^2}|(\alpha_kI^*_{r,k}+\beta^*_kI_{cr,k})|^2,
\end{split}
\end{equation}
where $\Phi(\textbf{x},t)=\int dk  [a_k\phi(\textbf{x},t)+a_k^{\dagger}\phi^*(\textbf{x},t)]$, $I_{cr,k}=\int_{-\infty}^{\infty}d\tau e^{i\Omega \tau}\phi^*(\textbf{x},t)$, and $I_{r,k}=\int_{-\infty}^{\infty}d\tau e^{-i\Omega \tau}\phi^*(\textbf{x},t)$. If we concentrate on the probability of exciting the atom, and integrate $\Gamma'_{\beta_k,k}$ over $\beta_k$, we obtain the total stimulated emission rate in mode $k$ as 
\begin{eqnarray}
\Gamma^{\alpha}_{cr,k}&=&\frac{1}{\pi T_t}\int d^2\beta_k |g\int d\tau\bra{\beta_k,e}\Phi(\textbf{x},t)\sigma_x(\tau)\ket{\alpha_k,g}|^2  \nonumber\\
&=&\frac{g^2}{\pi T_t}\int d^2\beta_k e^{-|\alpha_k-\beta_k|^2}|\beta_k^*I_{cr,k}+\alpha_k I^{*}_{r,k}|^2.
\end{eqnarray}
The massless scalar field $\Phi(\textbf{x},t)$ inside the cavity can be rewritten as
\begin{equation}
\begin{split}
	\Phi(\textbf{x},t)=\int_{-\infty}^{\infty}dk  [\phi_k(\textbf{x},t)a_k+H.c.],
	%=&\frac{1}{2\pi R}\sum_{m=-\infty}^{\infty}\sum_{n=1}^{\infty}\frac{B_{m}(\xi_{mn}r/R)}{B_{|m|+1}(\xi_{mn})}\int_{-\infty}^{\infty}\frac{dk_z}{\sqrt{\omega_k}} *(a_ke^{-i\omega_kt+im\theta+ik_zz}+H.c.),
\end{split}		
\end{equation}
where the eigen-function of the mode $k$ is given by 
\begin{equation}
\phi_k(\textbf{x},t)=\frac{1}{2\pi R}\frac{B_{p}(\xi_{pq}r/R)}{B_{|p|+1}(\xi_{pq})}\frac{1}{\sqrt{\omega_k}}\\
e^{-i\omega_kt+ip\theta+ik_zz}.
\end{equation}
For an atom undergoing uniform circular motion around the axis inside the cavity, we get $I_{cr.k}$ for counter-rotating coupling as
\begin{eqnarray}
I_{cr,k}&=&\frac{1}{2\pi R}\frac{B_{p}(\xi_{pq}r/R)}{B_{|p|+1}(\xi_{pq})}\frac{1}{\sqrt{\omega_k}}\int_{-\infty}^{\infty}d\tau e^{i\Omega \tau+i\omega_k\gamma\tau-ip\gamma\omega\tau}  \nonumber\\
&=&\frac{1}{2\pi R}\frac{B_{p}(\xi_{pq}r/R)}{B_{|p|+1}(\xi_{pq})}\frac{\delta(\Omega+\omega_k\gamma-p\gamma\omega)}{\sqrt{\omega_k}}.
\end{eqnarray}
Similarly for rotating-wave terms, we have
\begin{equation}
\begin{split}
	I_{r,k}=\frac{1}{2\pi R}\frac{B_{p}(\xi_{pq}r/R)}{B_{|p|+1}(\xi_{pq})}\frac{\delta(-\Omega+\omega_k\gamma-p\gamma\omega)}{\sqrt{\omega_k}}.
\end{split}
\end{equation}
Hence we can fixed the angular velocity $\omega$ of the atom and the frequency $\omega_k$ of the driving beams around the resonant point $\Omega+\omega_k\gamma-p\gamma\omega=0$. This ensures that resonant coupling $-\Omega+\omega_k\gamma-p\gamma\omega=0$ for rotation terms  cannot be satisfied in the same time. In this case we have $|I_{cr,k}|^2\gg|I_{r,k}|^2 \sim 0$ and
\begin{equation}
\begin{split}
	\Gamma^{\alpha}_{cr,k}&\simeq\frac{g^2}{\pi T_t}\int d^2\beta_k e^{-|\alpha_k-\beta_k|^2}|\beta_k^*I_{cr,k}|^2\\
	&=(|\alpha_k|^2+1)\Gamma_{cr,k}
\end{split}
\end{equation}
with $\frac{g^2}{T_t}|I_{cr(r),k}|^2=\Gamma_{cr(r),k}$ denoting the spontaneous emission rate in mode $k$ for counter-rotating coupling.

\section{Stimulated	emission rate for Dicke state interacted with a coherent input}
In Fig.3, $\Gamma^{\alpha,D}_{cr}$ and $\Gamma^{\alpha,D}_{r}$ are the total emission rates of $10^6$ atoms at superradiance state $\ket{\psi(t_0)}$ interacting with the coherent state $|\alpha_k|^2=10000$ for counter-rotating and rotating-wave couplings, respectively. For simplicity, we set the state of atoms as Dicke state $\ket{j,m}$(we can get the analogous result for $\ket{\psi(t_0)}$). As for rotating terms, since the resonant condition $-\Omega+\omega_k\gamma-p\gamma\omega=0$ is not satisfied, the effect of stimulated laser beams on the emission rate is negligible, and we can obtain the $\Gamma_{r}^{\alpha,D}$ by replacing the input with vacuum state as
\begin{eqnarray}
\Gamma^{\alpha,D}_r& \simeq &\frac{1}{T_t}\int dk\left|\int d\tau \bra{1_k}\bra{j,m-1}H_{int}\ket{0}\ket{j,m}\right|^2  \nonumber\\
&=&\frac{(j+m)(j-m+1)}{T_t}\int dk\left|\int d\tau \bra{1_k,g}H_{int}\ket{0,e}\right|^2  \nonumber\\
&=&(j+m)(j-m+1)\Gamma_{r}.	
\label{eq: P_R(k)}
\end{eqnarray}
Therefore $\Gamma_{r}^{\alpha,D}$ is enhanced by Dicke-like superradiance when $m=0$. As for counter-rotating terms, we obtain $\Gamma_{cr,k}^{\alpha,D}$ in mode $k$ as
\begin{eqnarray}
\Gamma_{cr,k}^{\alpha,D}&=&\frac{1}{\pi T_t}\int d^2\beta_k \left|\int d\tau \bra{\beta_k}\bra{j,m+1}H_{int}\ket{\alpha_k}\ket{j,m}\right|^2  \nonumber\\
&=&\frac{(j-m)(j+m+1)}{\pi T_t}\int d^2\beta_k \left|\int d\tau \bra{\beta_k,e}H_{int}\ket{\alpha_k,g}\right|^2  \nonumber\\
&=&(j-m)(j+m+1)(|\alpha_k|^2+1)\Gamma_{cr,k}  \nonumber\\
&=&(|\alpha_k|^2+1)\Gamma_{cr,k}^{D},
\label{eq: P_R(k)}
\end{eqnarray}
which is enhanced by both input coherent state and Dicke-like superradiance when $m=0$.  The total emission rate $\Gamma_{cr}^{\alpha,D}$ contains both spontaneous component and stimulated component. Thus we have 
\begin{eqnarray}
\Gamma_{cr}^{\alpha,D}&=&\sum_{k'\neq k}\Gamma^D_{cr,k'}+\Gamma^{\alpha,D}_{cr,k}  \nonumber\\
&=&\sum_{k'\neq k}\Gamma^D_{cr,k'}+(|\alpha_{k}|^2+1)\Gamma^{D}_{cr,k}  \nonumber\\
&=&\sum_{k'}\Gamma^D_{cr,k'}+|\alpha_{k}|^2\Gamma^{D}_{cr,k}
\end{eqnarray}
where the first term in the last line is the component related to spontaneous emission rate, and the second term is stimulated emission rate. Numerically, we find $ |\alpha_{k}|^2\Gamma^{D}_{cr,k}\gg\sum_{k'}\Gamma^D_{cr,k'}$ as $|\alpha_{k}|=10000$. Physically, the coherent light is so strong that we can ignore the contribution from spontaneous emission and only consider the component of stimulated emission, which thus indicates $\Gamma_{cr}^{\alpha,D} \simeq|\alpha_{k}|^2\Gamma^{D}_{cr,k}$.

\section{\label{A} The spontaneous emission rates of a spin coherent state and $\ket{\psi(t_0)}$ }
In this section, we present the calculation details for the spontaneous emission rates of a spin coherent state $|\theta,\phi\rangle=(\cos(\frac{\theta}{2}\ket{g}+e^{i\phi}\sin\frac{\theta}{2}\ket{e})^{\otimes N}$, and argue that the emission rate of $|\psi(t_0)\rangle$ can also be enhanced by superradiance during the evolution.

For a spin coherent state $|\theta,\phi\rangle$, the spontaneous emission rate for rotating-wave terms can be expressed as 
\begin{eqnarray}
\Gamma^{sc}_{r}&=&\sum_{m}|\bra{N/2,m}J_-\ket{\theta,\phi}|^2\Gamma_{r} =\bra{\theta,\phi}J_+J_-\ket{\theta,\phi}\Gamma_{r}  \nonumber \\
%&=(cos(\frac{\theta}{2}\bra{g}+e^{-i\phi}sin\frac{\theta}{2}\bra{e})^{\otimes N}\sum_{i=1}^{N}\sigma^{i}_{+}\\
%&\sum_{i=1}^{N}\sigma^{i}_{-}(cos(\frac{\theta}{2}\ket{g}+e^{i\phi}sin\frac{\theta}{2}\ket{e})^{\otimes N}\Gamma_{r}\\
&=&\left [C_{N}^{1}(C_{N}^{1}-1)\cos^2\frac{\theta}{2}\sin^2\frac{\theta}{2}+C_{N}^{1}\sin^2\frac{\theta}{2} \right ]\Gamma_{r}  \nn\\
&=&(\frac{N^2}{4}\sin^2\theta+N\sin^4\frac{\theta}{2})\Gamma_{r}
\end{eqnarray}
% \begin{equation}
%               \Gamma^{sc}_{r}&=\sum_{m}|\bra{N/2,m}J_-\ket{\theta,\phi}|^2\Gamma_{r} =\bra{\theta,\phi}J_+J_-\ket{\theta,\phi}\Gamma_{r}\\
%                  %&=(cos(\frac{\theta}{2}\bra{g}+e^{-i\phi}sin\frac{\theta}{2}\bra{e})^{\otimes N}\sum_{i=1}^{N}\sigma^{i}_{+}\\
%                  %&\sum_{i=1}^{N}\sigma^{i}_{-}(cos(\frac{\theta}{2}\ket{g}+e^{i\phi}sin\frac{\theta}{2}\ket{e})^{\otimes N}\Gamma_{r}\\
%                  &=\left [C_{N}^{1}(C_{N}^{1}-1)\cos^2\frac{\theta}{2}\sin^2\frac{\theta}{2}+C_{N}^{1}\sin^2\frac{\theta}{2} \right ]\Gamma_{r}\\
%                  &=(\frac{N^2}{4}\sin^2\theta+N\sin^4\frac{\theta}{2})\Gamma_{r}       
% \end{equation}
where $C_{n}^{m}=\frac{n!}{m!(n-m)!}$. In the same way, the spontaneous emission rate for counter-rotating terms can be expressed as 
\begin{equation}
\Gamma^{sc}_{cr}=(\frac{N^2}{4}\sin^2\theta+N\cos^4\frac{\theta}{2})\Gamma_{cr}.
\end{equation}
Therefore, superradiance occurs when $\theta=\pi/2$.

As for $|\psi(t)\rangle = \exp[-iH(t)] |\psi(0)\rangle $,  which is neither a superradiant Dicke state $\ket{j,m}$ with 
$m=0$ nor a spin coherent state with $\theta=\pi/2$,  it is necessary to prove that the emission rate induced by $\ket{\psi(t)}$ can also be enhanced by superradiance. Generally the spontaneous emission rate for counter-rotating terms of $\ket{\psi}$ can be expressed as
\begin{eqnarray}
\Gamma^{\ket{\psi}}_{cr}&=&\sum_{m}|\bra{N/2,m}J_+\ket{\psi}|^2\Gamma_{cr}=\bra{\psi}J_-J_+\ket{\psi}\Gamma_{cr}  \nn\\
&=&\bra{\psi}J_- \left(\ket{\psi}\bra{\psi} +\sum_{\ket{\psi_{\perp}}}\ket{\psi_{\perp}}\bra{\psi_{\perp}} \right) J_+\ket{\psi}\Gamma_{cr}  \nn\\
&=&\left(\avg(J_-)\avg(J_+)+\sum_{\ket{\psi_{\perp}}}|\bra{\psi_{\perp}}J_+\ket{\psi}|^2 \right)\Gamma_{cr} \nn\\
&>&\avg(J_-)\avg(J_+)\Gamma_{cr}
\end{eqnarray}
where $\avg(J_-)=\bra{\psi}J_-\ket{\psi}$, $\avg(J_+)=\bra{\psi}J_+\ket{\psi}$. As shown in Ref.\cite{PhysRevApplied.18.064006}, the mean value of $J_-(t)$ for the initial state $|\psi(0)\rangle =\ket{\frac{N}{2},\frac{N}{2}}$ is given by 
\bea
\langle J_-(t) \rangle \simeq -\frac{N}{2} \sin(\Omega_Rt)e^{-i(\omega_at-\theta)}.
\eea
Therefore, the emission rate at $\Omega_R t_0=\frac{\pi}{2}$ is expressed as
\begin{equation}
\Gamma^{\ket{\psi(t_0)}}_{cr}>\frac{N^2}{4}\Gamma_{cr},
\end{equation}
which is increased $N^2/4$ times due to superradiance.

\section{ \label{B} Equilibrium state of atoms and negative temperature. }
For $N$ two-level atoms inside a cylindrical cavity,  since the spontaneous emission related to counter-rotating and rotating-wave couplings  exists simultaneously, the population of the atoms in different levels usually changes over time $t$.  Meanwhile, there is also an equilibrium state in the system, where the changes in atomic population caused by these two different emissions cancel each other out. 

\begin{figure}[t]
\includegraphics[width=0.45\textwidth]{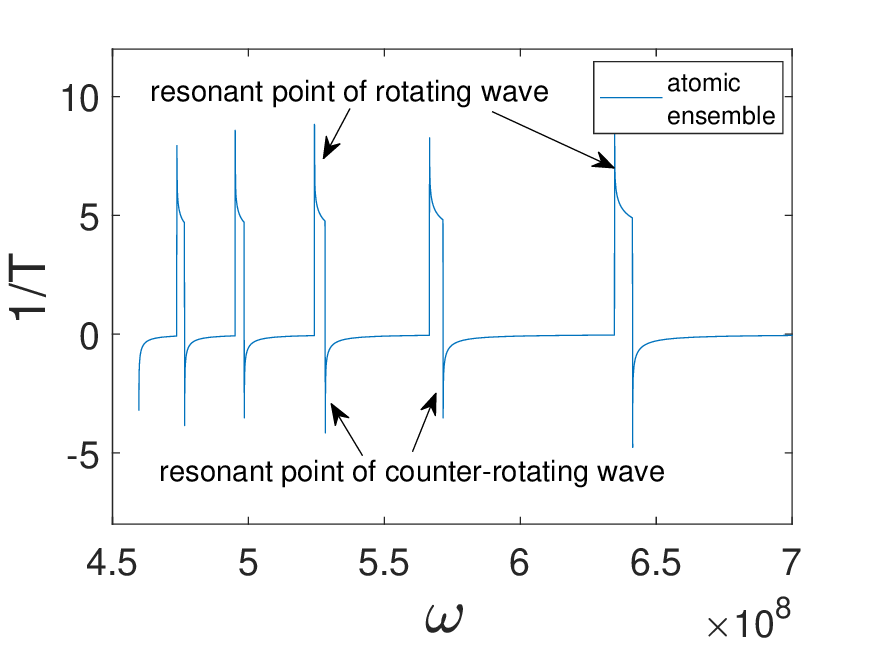}
\caption{The temperature $T$ as a function of the angular velocity $\omega$ of the atoms with the orbital radius $r=0.1{\rm m}$, $\Omega=10^7{\rm Hz}$. Other parameters are as follows: $R=1{\rm m}, g=10{\rm kHz}$.}
\label{fig4}
\end{figure}

For the Dicke-like model discussed above, the density matrix of the atomic ensembles can be expressed as $\rho=\sum_{m}P_m\ket{j,m}\bra{j,m}$,  with $\sum_{m}P_m=1$. As for the equilibrium state, $P_m$ can be determined by the detailed balance condition as
\bea
P_m\Gamma_r(j-m+1)(j+m)=P_{m-1}\Gamma_{cr}(j-m+1)(j+m).
\eea
The temperature $T$ of the ensemble can be defined by 
\begin{equation}
\frac{P_m}{P_{m-1}}=\frac{\Gamma_{cr}}{\Gamma_{r}}=e^{-\beta \Omega}
\end{equation} 
where $\beta=1/T$. The equilibrium temperature $T$ can also be derived directly by Lindblad master equation,  as shown in \cite{PhysRevA.70.012112}, where  the atoms are viewed as the system and the fields act as the environment, respectively.  Moreover, the definition of the temperature $e^{-\beta \Omega}=\Gamma_{cr}/\Gamma_{r}$ has also been used in \cite{PhysRevD.102.085006} for circular motion. Since both $\Gamma_{cr}$ and $\Gamma_{r}$ are functions of angular velocity, we can find the temperature of atomic ensemble as a function of $\omega$, which is plotted in Fig. (\ref{fig4}).

In Fig. (\ref{fig4}), the top and down peaks represent the resonant points of the rotating-wave and counter-rotating couplings, respectively. It should be noted that the temperature can be negative, which means the spontaneous emission rate of the counter-rotating coupling is larger than that induced by the normal rotating-wave coupling. That is very different from the cases with linear acceleration\cite{Ben_Benjamin_2019}, or with circular acceleration in free space without borders\cite{PhysRevD.102.085006}. Moreover, in the equilibrium case,  the atomic ensemble can emit photons continuously, while the state or  the temperature of the ensemble does not change.

\vspace{2cm}
\bibliography{referenceV}% Produces the bibliography via BibTeX.
\end{document}